\documentclass{cernyrep}

\usepackage[english]{babel}
\usepackage{verbatim}
\usepackage{graphicx}
\usepackage{epsfig}
\usepackage{amsfonts}
\usepackage{amsmath,amssymb}
\usepackage{bm}
\usepackage{hyperref}
\usepackage{color}
\usepackage{comment}
\usepackage[sort&compress,numbers]{natbib}
\usepackage{doi}

\usepackage{texnames}
\usepackage[T1]{fontenc}

\begin{document}

\title{The No-Core Gamow Shell Model: Including the continuum in the NCSM}
 
\author{B. R. Barrett$^{1}$, G. Papadimitriou$^{2,}$ \footnote{\href{georgios@iastate.edu}{georgios@iastate.edu}}, N. Michel$^{3}$, and M. P\l{}oszajczak$^{3}$}

\institute{
 $^{1}$ Department of Physics, University of Arizona, Tucson, Arizona 85721, USA\\
 $^{2}$ Department of Physics and Astronomy, Iowa State University, Ames, Iowa 50011-3160, USA\\
 $^{3}$ Grand Acc\'{e}l\'{e}rateur National d'Ions Lourds (GANIL),
CEA/DSM - CNRS/IN2P3, BP 55027, F-14076 Caen Cedex, France\\
}

\maketitle 
 
\begin{abstract}
We are witnessing an era of intense experimental efforts that will
provide information about the properties of nuclei far from the
valley of $\beta$ stability, regarding resonant and scattering states as well as
(weakly) bound states.  This talk describes our formalism for including these
 necessary ingredients into the No-Core Shell Model by using the Gamow
Shell Model approach.  Applications of this new approach, known as
the No-Core Gamow Shell Model, both to benchmark cases as well as
to unstable nuclei will be given.
\end{abstract}
 
\section{Introduction}

The atomic nucleus, which at energies in the range of mega-electron volts can be viewed as a quantum system of strongly interacting protons and neutrons  \cite{scidac,furnstahl,nuclei}, is a very fascinating object. One of the most intriguing challenges is the description of its time evolution. Indeed the nucleus can be naturally found in quantum states that decay by emitting photons, electrons or even nucleons and heavier particles. 
This endeavor becomes even more necessary in the rise of new experiments that probe new modes of radioactivity which will need theoretical justification \cite{mona}. 
One way is to  solve the time-dependent Schr\"{o}dinger equation by means of time discretization techniques  \cite{Bulgac10062011,volya,oishi}. A different approach  lies in  the solution of the time-independent Schr\"{o}dinger equation in the complex energy or momentum space. The time dependence is then absorbed by the complex nature of the solution, whose imaginary part is associated with the decay time. Metastable nuclear states and resonances could then be described in a time-independent formalism \cite{Gamow,Berggren,Romo,phil_trans}. 
Formulating the structure and reaction problem in the complex energy plane \cite{cs_real1,cs_real2,gaute_michel,Fossez, Laza,Laza2,Deltuva,Deltuva2,kruppa_scatcs,Myo20141,kikuchi}, provides with an alternative step towards the unification of structure (bound states) and reaction (resonances) aspects in nuclear physics, which will lead in a more controlled and model independent evaluation of observables. Noted that there
already exists vivid progress on the \textit{ab-initio} description of structure and reactions on the real-axis by  Lawrence Livermore and TRIUMF groups \cite{rgm1,rgm2,rgm3},
by Los Alamos/Argonne groups \cite{nollett_react} and also on the lattice by \cite{dean_lee,briceno}.
 
\section{The NC(GSM) formalism}

One of the ways to obtain  complex energy solutions of a physical system is by turning  the Schr\"{o}dinger equation into an eigenvalue problem and diagonalizing the complex Hamiltonian matrix.  
The basic code that has been employed for the description of resonances, by diagonalizing a very large non-Hermitian complex symmetric Hamiltonian matrix, is the Gamow Shell Model (GSM) code \cite{michel_2002,NM2,review_GSM,ncgsm,eff_inter}. The orthonormal underlying basis, upon which the Hamiltonian matrix is diagonalized, is known as Berggren basis \cite{Berggren} which provides a symmetric description of bound states, resonances and scattering states. The Hamiltonian matrix is non-Hermitian and complex symmetric with complex eigenvalues. The matrix is sparse and a relatively small number of eigenvectors and eigenvalues is of interest. At this point there is no conceptual difference between GSM with a core and the NCGSM besides the underlying Hamiltonian. Indeed, in the GSM case
the Hamiltonian consists of the one-body kinetic energy, the one-body mean-field (either schematic or Hartree-Fock) and the residual NN interaction, whereas in the NCGSM case the one-body mean field is not present in the A-body Hamiltonian \cite{ncgsm}. The nucleon-nucleon interaction is expressed in the Berggren basis using the potential separable expansion (PSE) \cite{gyar_kruppa} in a HO basis method \cite{hagen_morten_michel,gsm_radii,ncgsm}. In this way, matrix elements between scattering and/or resonant states never diverge, due to the Gaussian fall-off of the HO radial form factor, and in addition one can conveniently transform matrix elements from the 
relative frame to the lab frame. For matrix elements of other operators, such as electromagnetic transition operators,  the renormalization of integrals relies on the method of external complex scaling (ECS). 
We note here though that the PSE method has also been employed  for the calculation of the recoil matrix elements of the intrinsic Hamiltonian, since in this case the ECS technique does not provide converged
results.

For the diagonalization of the large complex symmetric matrix we have used 
a complex extension of the Lanczos algorithm \cite{review_GSM}. 
The largest matrix that has been diagonalized has a dimension of $\sim$ 10$^6$. This number materializes to about A=4,5 for \textit{ab-initio} no-core calculations in the Berggren basis and to 7-8 active valence particles when assuming a frozen configuration (also known as ``core") but allowing only a portion of the particles occupying the continuum (Berggren) orbitals (particle-hole truncations).
The second alternative is the Davidson method \cite{review_GSM} (see also a recent application on the description of tunneling for a two-body atomic system \cite{Forssen}).    
 We would like to highlight a unique feature of the GSM diagonalization. Both the Lanczos and the Davidson methods calculate the ground state of the system as the lowest eigenvalue. In the GSM where the Hamiltonian matrix is non-Hermitian, the lowest eigenvalue is not guaranteed to coincide with the ground state and it may as well be a scattering state. A criterion is established which separates the state of interest from the wealth of scattering states; that is the overlap method \cite{review_GSM}.
In the overlap method a smaller diagonalization takes place first, in a space spanned by
\begin{table}[h!]
\caption{Comparisons \cite{Pap_Maris} between the NCGSM and the computer program MFDn. N$_{\rm{shell}}$ = 2n+$\ell$  and $\hbar$$\omega$ = 20 MeV. }
\centering
\begin{tabular}{c c c c}
\hline\hline
Nucleus & MFDn & NCGSM & Difference \\ [0.5ex] 
\hline
$^2$H 1$^+$ (N$_{\rm{shell}}$ = 4)  & -1.6284  & -1.6284  & $\leq$ 0.1 keV  \\
$^2$H 1$^+$ (N$_{\rm{shell}}$ = 8)  & -2.1419    & -2.1419   & $\leq$ 0.1 keV \\
$^3$H 1/2$^+$ (N$_{\rm{shell}}$ = 4)  &  -7.6016   &   -7.6016   & $\leq$  0.1 keV \\
$^3$H 1/2$^+$ (N$_{\rm{shell}}$ = 8)  &  -8.3203   &   -8.3203   &   $\leq$ 0.1 keV \\
$^3$He 1/2$^+$(N$_{\rm{shell}}$ = 8)  &   -7.6084  & -7.6084 & $\leq$ 0.1 keV\\
$^4$He 0$^+$ (N$_{\rm{shell}}$ = 4) &     -27.3685 &  -27.3684 &   0.1 keV \\
$^6$Li 1$^+$ (N$_{\rm{shell}}$ = 4) & -24.9778 & -24.9776 &    0.2 keV \\ [1ex]
$^6$Li 3$^+$ (N$_{\rm{shell}}$ = 4) & -22.4959 & -22.4957 &   0.2 keV \\ [1ex]
\hline
\end{tabular}
\label{tab:1}
\end{table}
a few states (usually single particle bound states and resonances). This smaller space sometimes is also called the ``pole approximation". The solution is the reference eigenvector.
At a second step a diagonalization in the full space (bound states+resonances+scattering states) takes place and the solution is the one that maximizes the overlap with the reference eigenvector. 
    
Together with the GSM, another algorithm known as the Density Matrix Renormalization Group (DMRG) has also been used \cite{White,Papenbrock_dmrg,Pittel,Rotureau_2006,ncgsm}. The method is a truncation algorithm which aims in reducing the sizes of the matrices to be diagonalized, while keeping the same accuracy as in the full calculation. DMRG is an iterative method for which at each step the space is increased by adding basis states one-by-one and the truncation is dictated by
the density matrix which is constructed in each step. At each step a complex symmetric non-Hermitian matrix is diagonalized via the Lanczos method and the matrix is smaller than the typical matrix of a full-fledged GSM calculation. Several diagonalizations of such smaller matrices are performed until a convergence criterion is reached.  A recent variant of the DMRG method \cite{Legeza:2015fja} has shown that very large model spaces can be  reached and many ``sweeps" can be performed in a timely manner, making it tempting to use such a variant also in Berggren basis DMRG based calculations.
  
\section{Applications of the NCGSM approach}

\subsection{Benchmarks}
Aiming at a predictive theory it becomes increasingly important to complete a quality control on the solvers which are employed for the solution of
the many-body problem, a task that implies validation and cross-checks (benchmarks) of existing codes \cite{witek_jacek_stat}.
To test the NCGSM algorithm we have performed benchmark 
calculations, in which we compared the NCGSM results with
 results obtained using the NCSM \cite{Barrett2013131}  Many-Fermion Dynamics nuclear (MFDn)  computer program \cite{mfdn,mfdn2},
 and in Table (\ref{tab:1}) we present applications for systems up to $^6$Li. For this benchmark we employed a Harmonic Oscillator (HO) basis
 in a Full Configuration Interaction (FCI) truncation and
the JISP16 realistic interaction \cite{jisp16}. Nevertheless, chiral interactions were also tested successfully \cite{Pap_Maris}. This agreement reassures that the calculations are
not contaminated with unintentional errors or flaws. It should be noted that there has also been two other successful works that benchmarked the GSM algorithm, using central interactions and a $\alpha$-core,
against the complex scaling technique \cite{kruppa_george,masui}.

\subsection{ANCs and widths}

In its current implementation the NCGSM is not ready to provide  scattering observables on the real-axis, such as cross-sections, even though such 
a goal is not far from reach after the combination of the GSM with the Resonating Group Method \cite{rgm_book} using phenomenological interactions/optical potentials \cite{yannen,Fossez}.
\begin{figure}[h!] 
 \centering
  \includegraphics[width=100mm]{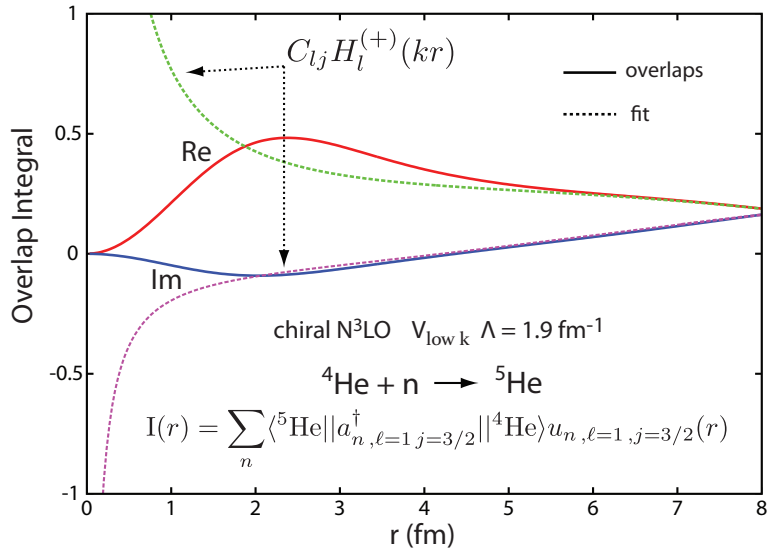} 
  \caption[T]{\label{Fig1}
  (Color online) Overlap function and tail fit with a Hankel function. C$_{lj}$ stands for the ANC. Figure is from \cite{ncgsm}.}
\end{figure}
We are able however to compute overlaps between nuclear states and access information associated with the ``tail" of the overlap.
The relevant quantity in the study of asymptotic properties of the nuclear wavefunction or actually its projection onto cluster (sub-cluster) states \cite{nollett_2011}
is the Asymptotic Normalization Coefficient (ANC).  Recently there is a collective effort in nuclear theory to compute asymptotic quantities and
we are witnessing  an abandonment of quantities such as spectroscopic factors in favor of ANCs and widths or partial widths.
The basic argument behind this endeavor, besides the physics interest (e.g. relevance to astrophysical processes for both resonance widths and ANCs),
 is the fact that asymptotic quantities are less model dependent and closer to the notion of an observable quantity \cite{kadyrov,furn_achim}. At the same time ANCs can serve as an internal consistency test between many-body methods since
 calculations at distances far away from the nuclear interaction range always pose challenges and difficulties (see e.g. discussion at \cite{nollett_2012} for some of the \textit{ab-initio} methods).
The GSM or the NCGSM which are formulated on a basis that has a correct asymptotic behavior and captures the relevant long range physics, become
 appropriate for the calculation of asymptotic quantities. For a detailed review of the progress that has been made in the calculation of ANCs and also the experimental 
 situation we refer the reader to \cite{nollett_2012} and also \cite{okolowicz} for ANCs calculations within the GSM. 
 
 In this contribution we  present calculations that were published in \cite{ncgsm} of ANCs within the NCGSM using realistic interactions.
 The model space of the calculation, as it was described in \cite{ncgsm}, includes single particle (s.p.) partial waves with angular momentum up to $\ell$ = 4 (g-waves).
 For $^5$He, being particle unstable in its ground state, we employed a complex GHF basis consisting of the s.p. 0p$_{3/2}$ resonant state and non-resonant p$_{3/2}$
 states along the complex contour which encloses the s.p. resonant state, a necessary condition for the s.p. Berggren completeness to be satisfied.
 In order to obtain the many-body solution and calculate the overlap for the reaction: $^4\rm{He}_{0_+}$ + n $\to$ $^5\rm{He}_{3/2^-}$ we used the Davidson diagonalization \cite{review_GSM}  method and limited our selves to continuum configurations that allowed up to four particles occupying continuum orbits (4p-4h truncation). In the future it will be important to assess the importance of the configurations involving many particles in continuum orbits and also try to
 accompany the results with a truncation error associated with the missing configurations. It should also be noted that the creation operator for the calculation of the overlap does not only
create on a single state,
 but there is a sum over all continuum states; a fundamental difference between NC(GSM) and traditional configuration interaction calculations in a HO basis \cite{gsm_SF}.

 We present our results in Fig.\ref{Fig1} which was taken from \cite{ncgsm} for the effective V$_{low\, k}$ \cite{Bogner20031,HjorthJensen1995125,cens}  $\Lambda$ = 1.9fm$^{-1}$ chiral N$^3$LO NN interaction \cite{EM_pot}.
 We see that the overlap exhibits both real and imaginary parts reflecting the complex nature of the Berggren basis. After fitting the asymptotic part of the overlap 
 with a complex Hankel function we extract an ANC with a real part of 0.197 fm$^{-1/2}$. Now knowing the ANC we are able to obtain the width of $^5$He using the formula:
 \begin{equation}
 \label{anc_width}
C = \sqrt{ \frac{\Gamma \mu}{\hbar^2 \Re(k)}  }
\end{equation}
 that relates the width and the ANC \cite{okolowicz}, where $\mu$ is the reduced mass, $k$ is the real part of the complex linear momentum that corresponds to the 
 neutron-separation energy of $^5$He and $C$ stands for the ANC. The result for the width is $\Gamma$ = 311 keV.
 Within the 4p-4h truncation, the complex Davidson
 diagonalization provided for the S$_{1n}$ the value of -1.561 MeV and a width $\Gamma$ of 370 keV. The small difference between the width obtained from the ANC
 formula and the one obtained from the diagonalization stems from an approximation that was made on the proof of formula \eqref{anc_width}.  The approximation is that
 the real part of the linear momentum has to be considered. It has been shown in \cite{ncgsm} that this approximation implies the condition $\frac{\Gamma}{2S_{1n}} \to $ 0,
 namely, formula \eqref{anc_width} will work for states that have
widths much smaller as compared to two-times the separation energy. In our case $\frac{\Gamma}{2S_{1n}}$  = 11.8 $\%$, 
 which  explains the deviation from the exact diagonalization result.
 
 
\subsection{NCGSM results for $^{4}$H and $^4$Li}
\begin{figure}[h!] 
 \centering
  \includegraphics[width=100mm]{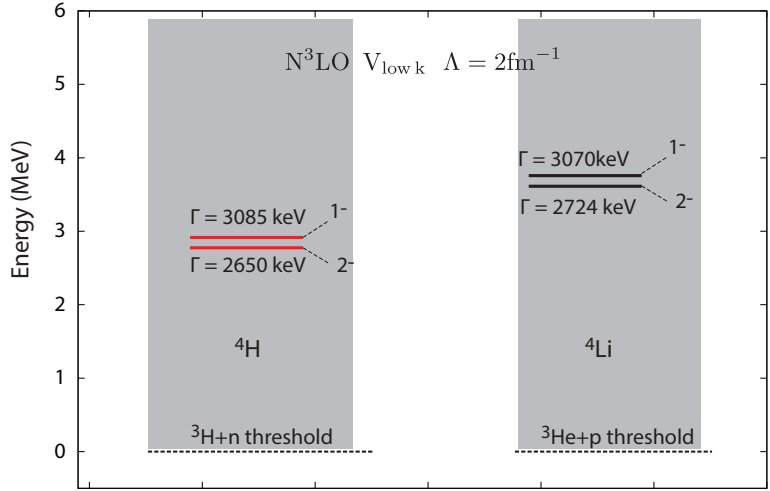} 
  \caption[T]{\label{Fig2}
  (Color online) Ground state and first excited state spectra of $^4$H and $^4$Li. Gray scale denotes the continuum/scattering regime which is described by adopting a complex Beggren basis.}
\end{figure}
Next we are presenting calculations for $^4$H and $^4$Li. These systems have also been computed in approaches with realistic interactions,
in \cite{scat_benchm} for a scattering benchmark calculation within several few body methods and in \cite{horiuchi} within the complex scaling method. 
Especially for hydrogen isotopes there exists a recent experimental interest since measurements
claim to observe a relatively narrow $^7$H resonance \cite{caamano}. In addition, it is believed that the triplet of isotopes $^{5,6,7}$H resemble the Helium isotopic chain ($^{6,7,8}$He). Namely
 $^7$H resembling $^8$He is the most ``bound" member of the triplet having a relatively small width, $^6$H resembling $^7$He is ``unbound" reflecting its extremely large width while $^5$H is  just a broad resonance (broader than $^7$H), resembling in this case $^6$He which is less bound than $^8$He. The theoretical investigation of the shell structure and the pairing correlations in the continuum
 will be a decisive factor in the understanding of the binding mechanism in this area of the chart and it 
will also be a challenging task. 
 
In this work we are using a phenomenological WS potential for the generation of the basis. For $^4$H and $^4$Li  the
WS basis is created for the $^3$H + n or $^3$He + p systems respectively.
Namely for $^4$H a neutron 0$p_{3/2}$ s.p. resonant state is considered
whereas for $^4$Li a proton 0$p_{3/2}$ s.p. resonant state. The long-range Coulomb interaction in the case of $^4$Li is treated in the same way as in \cite{ncgsm,michel_coul,coul_isosp}.
Besides the complex neutron and proton resonances and the associated complex continua, we consider s.p. partial waves up to h-waves ($\ell$=5) as HO basis states characterized by a frequency of $\hbar \omega$ = 20 MeV.  In particular, for $^4$H and for the neutron space we used a basis that consisted of 15p$_{3/2}$ complex s.p. states and  7s$_{1/2}$, 7p$_{1/2}$, 2d, 2f, 2g, 1h real HO s.p. states, while for the proton space all  basis states are HO states with 7s$_{1/2}$, 7p$_{1/2}$, 2d, 2f, 2g, 1h. 
Noted that we are using a different truncation scheme that departs from the typical N$_{max}$ or N$_{shell}$ HO truncation. Here, and also in other NCGSM calculations, we choose to use more radial nodes for specific $\ell$ states. Here states with $\ell$=0,1 have been chosen to have more nodes since
we noticed that they contribute more energy to the system. Further calculations that investigate this behavior are in progress.
The same basis was used for $^4$Li but instead of a neutron s.p. resonant state we considered a proton s.p. resonant state.
Our results for the g.s and first excited states of $^4$H and $^4$Li are shown on Fig.\ref{Fig2}. The energies are with respect to the one-neutron and one-proton particle thresholds.
For the thresholds, the g.s. energies of $^3$H and $^3$He are found to be -7.92 MeV and -7.12 MeV respectively, for an effective V$_{low\, k}$ $\Lambda$ = 2.0fm$^{-1}$ N$^3$LO Chiral EFT interaction.
Overall we observe a good agreement with experimental measurements \cite{tunl} and especially the small gap between the 1$^-$ and 2$^-$ states that has been observed is also predicted in our calculation.
Even though the calculations for the widths show stability, the widths are very large, so we are not dealing with typical resonances that would have an appreciable impact on cross-sections.
In the calculation presented here we restricted the occupation of continuum orbits to a maximum of three particles (3p-3h).
In our future work we will provide results without truncations and also with investigations on the impact of missing excitations on energies and widths.

\section{Conclusions and outlook}
In conclusion, we presented applications of the NCGSM for the calculation of energies, widths and asymptotic quantities such as ANCs of unbound nuclei. At the same
time we benchmarked our algorithm against another commonly used solver.
Our immediate goals are to continue working on applications of the NCGSM for light unbound nuclear systems in the neutron and proton rich side of the nuclear chart.
The success of the NCGSM is tied with advances in computer algorithms and the formulation of efficient complex symmetric diagonalization solvers. On the physics side, even 
when the most efficient solver will be at hand, a full calculation in the Berggren basis (no p-h truncations) for a system such as $^{11}$Li will be very demanding. Hence, one of our goals is
to quantify at first the impact of missing truncations (we already know that the weight of many particles in continuum orbits is small). Also we aim at a construction of a realistic effective interaction, i.e. an  interaction in the NCGSM nuclear medium \footnote{NCGSM medium here should be seen as the model space that the A-nucleons evolve, but in addition, scattering to continuum orbits will be allowed through the Berggren basis} that
would stem from a realistic free-space interaction, which we will then utilize for NCGSM calculations (see e.g. \cite{gsm_eff_gaute} for a Lee-Suzuki transformation for interactions in the complex energy plane and a multireference perturbation theory approach).
Finally, we would like to contribute to the effort of bridging the gap between Lattice-QCD and low energy physics and a possible way was shown in \cite{bira_lat}. This could be achieved once NCGSM will be utilized to handle NNN interactions. Then we could naturally compute resonant features of finite nuclei at any pion mass
that is available at that time.

\subsection{Acknowledgments}
This work was supported by the US DOE under grants No. DESC0008485 (SciDAC/NUCLEI). This work was also supported in part by the FUSTIPEN (French-U.S. Theory
Institute for Physics with Exotic Nuclei) under U.S. DOE grant No. DE-FG02-10ER41700.
We would like to thanks H. M. Aktulga, W. Nazarewicz, J. P. Vary, C. Yang, E. G. Ng and G. I. Fann for fruitful discussions on complex energy diagonalization solvers
 and P. Maris for his help with the benchmark calculations of Table I.

\bibliographystyle{apsrev4-1}  
\bibliography{Varenna_bib}    

\end{document}